\documentclass{elsart}
\usepackage{graphicx}
\newcommand{\idxtxt}[1]{\mathrm{#1}} \newcommand{\ii}{\mathrm{i}}

\newcommand{\odn}[3]{\frac{d^{#3}#1}{d#2^{#3}}}
\newcommand{\od}[2]{\frac{d#1}{d#2}}

\newcommand{\pdn}[3]{\frac{\partial^{#3}#1}{\partial#2^{#3}}}
\newcommand{\pd}[2]{\frac{\partial#1}{\partial#2}}

\newcommand{\sign}{\operatorname{sign}}

\begin{document}
\begin{frontmatter}
\title{Nonlinear Bessel beams}

\author{Pontus Johannisson,} \author{Dan Anderson,} \author{Mietek
Lisak,} \author{Mattias Marklund}

\address{Dept. of Electromagnetics, Chalmers University of Technology,
SE--412~96~G\"oteborg, Sweden}

\begin{abstract}
The effect of the Kerr nonlinearity on linear non-diffractive Bessel
beams is investigated analytically and numerically using the nonlinear
Schr\"odinger equation. The nonlinearity is shown to primarily affect
the central parts of the Bessel beam, giving rise to radial compression
or decompression depending on whether the nonlinearity is focusing or
defocusing, respectively.  The dynamical properties of
Gaussian-truncated Bessel beams are also analysed in the presence of a
Kerr nonlinearity.  It is found that although a condition for width
balance in the root-mean-square sense exists, the beam profile becomes
strongly deformed during propagation and may exhibit the phenomena of
global and partial collapse.
\end{abstract}

\end{frontmatter}

\section{Introduction}
Diffractive spreading of waves is a classical phenomenon in wave
dynamics and an inherent feature of beam propagation.  Much attention
has been devoted to the possibility of counteracting the dispersive
spreading by focusing effects due to medium nonlinearities e.g.\ the
Kerr effect, cf.~\cite{marburger}.  However, it has also been pointed
out, \cite{durnin}, that non-diffracting beams are possible also in
linear media.  In particular, the Helmholz equation that governs the
linear diffractive dynamics of a wave beam allows classes of
diffraction-free solutions.  In addition to the plane wave solutions,
the two-dimensional counterparts, the cylindrically symmetric Bessel
solutions, also propagate with preserved form, while also allowing for
a concentrated beam profile.  The drawback from an application point
of view is the fact that these beams have infinite energy, and
consequently cannot be realized physically.  Various ways to
circumvent this problem have been suggested, the most obvious being to
truncate the Bessel beam at some radius e.g. by a Gaussian truncation,
forming the so called Bessel-Gauss beams, \cite{gori}.  While such a
truncation clearly reintroduces diffraction, the beam broadening could
be made small if the propagation length is kept smaller than the
corresponding diffraction length of the Bessel-Gauss beam.  In
particular, since the Bessel beam diffracts sequentially, starting
with the outer lobes, cf.~\cite{sprangle}, the central part of the
beam remains intact for a certain distance of propagation.

Recently, there has been growing interest in nonlinear effects in
connection with Bessel and Bessel-Gauss beams,
\cite{sogomonian,ding,gadonas,butkus}.  Of special interest for the
present investigation is the attention given to media with an
intensity dependent refractive index, i.e., Kerr media, see
e.g.~\cite{gadonas}.  The work carried out in \cite{gadonas} considers
the limit of weak nonlinearity, which makes it possible to use a
perturbation approach involving an expansion around the lowest order
linear (stationary) Bessel solution for solving the evolution
equation, being the nonlinear Schr\"odinger equation.

In the present paper, we investigate in more detail the nonlinear
generalisation of the linear diffraction-less Bessel beam solutions as
well as the nonlinear dynamics of Bessel-Gauss beams.  Stationary
solutions in two dimensions are determined by the Bessel equation
modified by a nonlinear term, i.e., the radially symmetric nonlinear
Schr\"odinger equation.  The modified Bessel solutions, the
``nonlinear Bessel beams'', are studied using approximate analytical
and numerical methods.  The results show that the nonlinearity
primarily affects the central high intensity parts of the beam
profile, which become radially compressed or decompressed depending on
whether the nonlinearity is focusing or defocusing, respectively.  The
beam profile for large radii remains a Bessel function with a phase
shift being the only remaining effect of the nonlinearity.  However,
for the defocusing nonlinearity an amplitude threshold exists, above
which no solutions decaying to zero exist.

The dynamical properties of Gaussian-truncated Bessel beams in the
presence of a Kerr nonlinearity are also studied.  An exact analytical
solution was previously found for the linear dynamics of the
Bessel-Gauss beams, \cite{gori}.  Based on the virial theorem, which
gives an exact analytical description of the variation of the beam
width in the root-mean-square (RMS) sense, important information about
the effect of the nonlinearity on the beam dynamics can be obtained.
In particular, it is found that a focusing nonlinearity tends to cause
an evolution stage where the central parts of the Bessel beams are
initially compressed.  Depending on the strength of the nonlinearity,
different scenarios are possible, e.g. the subsequent evolution may
involve an essentially diffraction-dominated behaviour, but for
increasing nonlinearity, two forms of collapse may appear.  Either a
part of the beam collapses, while the RMS width of the beam still
increases, or above a certain threshold, the RMS width goes to zero in
a finite distance.  Numerical simulations of the dynamics illustrate
the different scenarios.

\section{The  nonlinear  Schr\"odinger  equation}
The propagation of an optical wave in a nonlinear Kerr medium is
described by the nonlinear Schr\"odinger equation.  This implies that
the slowly varying wave envelope, $\psi(z,r)$, of a cylindrically
symmetric beam satisfies the following equation
\begin{equation}
\ii \pd{\psi}{z} = \frac{1}{2 k_0} \left( \pdn{\psi}{r}{2} +
\frac{1}{r} \pd{\psi}{r} \right) + \kappa |\psi|^2 \psi,
\label{eq_nlse}
\end{equation}
where $z$ is the distance of propagation, $k_0$ is the wave number in
vacuum, and $\kappa$ is the nonlinear parameter.  Additional physical
effects, e.g.\ attenuation and gain, can be modelled by using complex
coefficients in Eq.~(\ref{eq_nlse}).  The obtained complex equation,
which in one dimension has analytical soliton solutions,
\cite{pereira}, is the cylindrical generalisation of the
Pereira-Stenflo equation. It has been investigated using a variational
approach, \cite{anderson2}, but further work is needed to fully
describe the complex case.  In the present work, the coefficients are
assumed to be real.  It is convenient to introduce the normalisation
$\tilde{r} = r/a_0$, where $a_0$ is a characteristic width of the
beam, and $\tilde{z} = z/L_D$, where $L_D \equiv 2 k_0 a_0^2$ is the
Rayleigh length.  Eq.~(\ref{eq_nlse}) then takes the form
\begin{equation}
\ii \pd{\psi}{\tilde{z}} = \pdn{\psi}{\tilde{r}}{2} +
\frac{1}{\tilde{r}} \pd{\psi}{\tilde{r}} + \tilde{\kappa} |\psi|^2
\psi,
\label{eq_nlse_norm}
\end{equation}
where $\tilde{\kappa} = L_D \kappa$. For simplicity we will suppress
the tilde in the subsequent expressions.

We begin the analysis by looking for stationary solutions of
Eq.~(\ref{eq_nlse_norm}). For this purpose we write $\psi = \psi(z,r)
= A(r) \e^{\ii \delta z}$, which leads to the eigenvalue equation
\begin{equation}
\odn{\!A}{r}{2} + \frac{1}{r} \od{A}{r} + \delta A + \kappa A^3 = 0.
\label{eq_nonlin_bessel_eq}
\end{equation}
This equation is to be solved subject to the boundary conditions that
the solution should be finite when $r = 0$ and go to zero as
$r\rightarrow\infty$.  The lowest order solution in the physical
situation when the nonlinearity balances the diffraction, i.e., the
focusing case with $\kappa >0$, corresponds to the so called Townes
soliton~\cite{chiao}, which has essentially the same properties and
$sech$-shaped form as the one-dimensional soliton solution,
cf.~\cite{anderson}. In particular, this solution only exists for
negative eigenvalues, which are uniquely related to the maximum
amplitude.

The Bessel beams are the solutions of the linear Schr\"odinger
equation ($\kappa = 0$) and are given by $A(r) = A_0 J_0(\sqrt{\delta}
\, r)$. Clearly, well-behaved solutions exist only for positive
eigenvalues $\delta$. In contrast to the nonlinear case, the linear
eigenvalue problem has a continuous set of (positive) eigenvalues,
which are independent of the amplitude of the beam profile.  The first
task of the present analysis is to analyse the nonlinear Bessel beams,
being the solutions of Eq.~(\ref{eq_nonlin_bessel_eq}) for positive
$\delta$ and $\kappa \neq 0$.  We note that by introducing 
\begin{equation}
\bar{r} = \sqrt{|\delta|} \, r, \quad \bar{A} = \sqrt{|\kappa/\delta|}
\, A,
\end{equation}
only the signs of $\delta$ and $\kappa$ remain in
Eq.~(\ref{eq_nonlin_bessel_eq}).  Thus, without loss of generality, it
will be assumed that $\delta = 1$ and $\kappa = \pm 1$.

\section{Nonlinear Bessel beams}
In order to analyse the properties of the nonlinear Bessel beams
analytically it is instructive to start by examining the central part
of the pulse, which is determined by $\sqrt{\delta} \, r \ll
1$. Since the initial derivative of $A(r)$ must be zero, we have the
approximation $A(r) = A_0 + O(r^2)$, which implies that to second
order in $r$, Eq.~(\ref{eq_nonlin_bessel_eq}) can be approximated by
the linear equation
\begin{equation}
\odn{\!A}{r}{2} + \frac{1}{r} \od{A}{r} + (\delta + \kappa A_0^2) A =
0, \label{eq_nonlin_eq_sol_small_r}
\end{equation}
with the corresponding solution
\begin{equation}
A = A_0 J_0(\sqrt{\delta + \kappa A_0^2} \, r).
\end{equation}
This solution is valid for small $r$ only, but nevertheless gives
important information about the nonlinear modifications of the Bessel
beam. In the focusing case ($\kappa > 0$), the main lobe tends to be
compressed and we expect that the amplitude of the nonlinear Bessel
beam will oscillate more rapidly than the linear Bessel beam. On the
other hand, in the defocusing case ($\kappa < 0$), the main lobe
should become wider and the nonlinear solution should oscillate slower
than the linear one. In particular, if the amplitude is chosen to
fulfil
\begin{equation}
\delta + \kappa A_0^2 < 0, \label{eq_limit}
\end{equation}
the expression under the square root will be negative.  This
corresponds to the modified Bessel function, which is growing with
$r$.  Thus the presence of a defocusing nonlinearity can qualitatively
change the behaviour of the solution.

It is clear that if the second derivative of $A$ is positive
initially, it will remain positive.  This can be seen by rewriting the
equation as
\begin{equation}
\odn{\!A}{r}{2} + \frac{1}{r} \od{A}{r} + \delta_{\idxtxt{eff}} A = 0,
\end{equation}
where $\delta_{\idxtxt{eff}} \equiv \delta + \kappa A^2$.  If
$\delta_{\idxtxt{eff}}$ is negative at $r = 0$, the solution will be
growing for small $r$ and $\delta_{\idxtxt{eff}}$ will be further
decreased.  This increases the derivative of the solution, and implies
that a sufficiently strong defocusing nonlinear term will give rise to
a monotonically growing solution, which is not compatible with the
condition at infinity.

A more accurate description of the main lobe of the solution can be
obtained using variational analysis and Ritz optimisation,
cf.~\cite{anderson}. When using variational analysis, it is important
to find a good set of trial functions that gives tractable
calculations while maintaining sufficient accuracy.  A trial function
that should approximate the main lobe of the nonlinear Bessel beam
reasonably well is
\begin{equation}
A_T = A_0 J_0 \left( j_0 \frac{r}{r_0} \right), \label{eq_ansatz}
\end{equation}
where $j_0$ is the first zero of the Bessel function.  This also has
the advantage that the exact linear result is recovered in the limit
$\kappa \to 0$.  In the variational procedure, we assume $r_0$ to be
given and consider $A_0$ as a free parameter.  The Lagrangian
corresponding to Eq.~(\ref{eq_nonlin_bessel_eq}) is
\begin{equation}
\mathcal{L} \equiv \langle L \rangle = \int_0^{r_0} L[A_T] \d r,
\end{equation}
where
\begin{equation}
L[A] = \frac{r}{2} \left( \od{A}{r} \right)^2 - \frac{r \delta A^2}{2}
- \frac{r \kappa A^4}{4}.
\end{equation}
Using the ansatz (\ref{eq_ansatz}) we obtain
\begin{equation}
\mathcal{L} = \frac{1}{4} \left[ j_0^2 A_0^2 J_1^2(j_0) - r_0^2 \delta
A_0^2 J_1^2(j_0) - C_1 r_0^2 \kappa A_0^4 \right],
\end{equation}
where 
\begin{equation}
C_1 = \int_0^{1} J_0^4(j_0 x) x \d x \approx 7.62 \times 10^{-2}.
\end{equation}
The variation with respect to $A_0$, i.e., $\partial \mathcal{L}/
\partial A_0 = 0$ yields
\begin{equation}
A_0^2 = \frac{(j_0^2 - r_0^2 \delta) J_1^2(j_0)}{2 C_1 r_0^2 \kappa}
\quad \Longleftrightarrow \quad r_0^2 = \frac{j_0^2 J_1^2(j_0)}{\delta
J_1^2(j_0) + 2 C_1 \kappa A_0^2}.
\label{eq_var_result_amp}
\end{equation}
By setting $\kappa = 0$, the linear result $\sqrt{\delta} \, r_0 = j_0
$ is recovered. If $\kappa$ is positive the value of $r_0^2$ is
decreased, which corresponds to compression of the main lobe. A
negative $\kappa$ gives the opposite effect. This result confirms the
previously obtained picture of the effects of the nonlinearity on the
main lobe of the linear Bessel solution.

\begin{figure}[htbp]
\begin{center}
\includegraphics[scale=0.6]{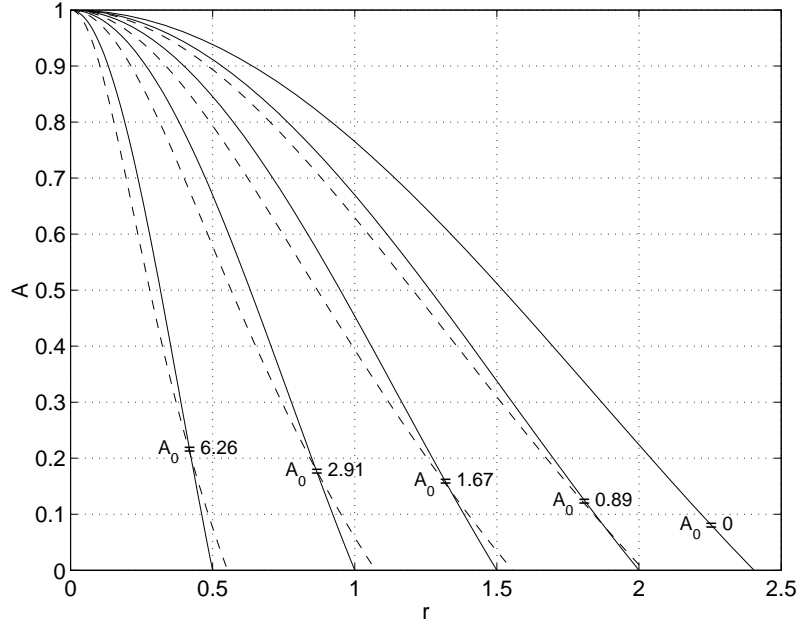}
\caption{Comparison of the variational results (solid lines) to the
numerical ones (dashed lines) for a focusing nonlinearity.}
\label{fig_var_num_cos_focus}
\end{center}
\end{figure}

\begin{figure}[htbp]
\begin{center}
\includegraphics[scale=0.6]{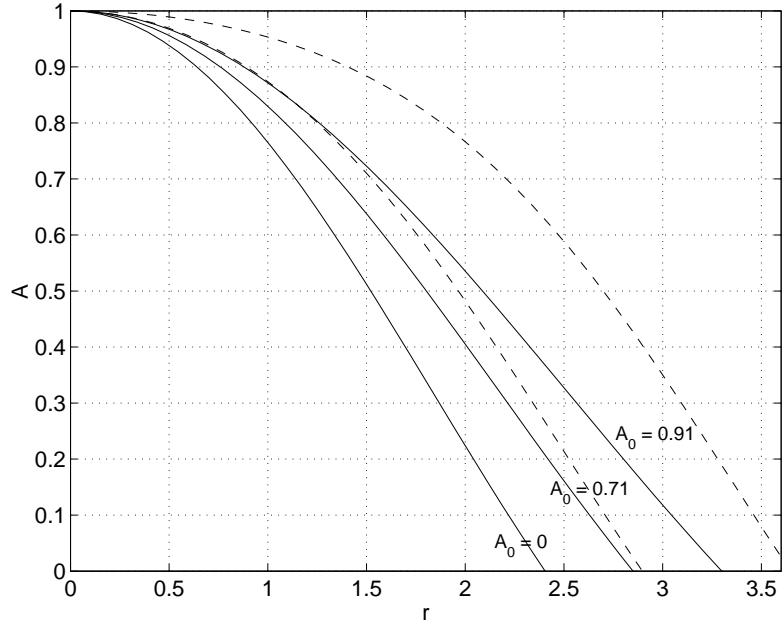}
\caption{Comparison of the variational results (solid lines) to the
numerical ones (dashed lines) for a defocusing nonlinearity.}
\label{fig_var_num_cos_defocus}
\end{center}
\end{figure}

The result of the variational analysis is compared with numerical
solutions in Figs.~\ref{fig_var_num_cos_focus} and
\ref{fig_var_num_cos_defocus}.  Different $r_0$-values have been used,
and they are easily identified in the figures, since the variational
approximation is zero when $r = r_0$.  For clarity the plotted curves
have been normalised with respect to their amplitudes at $r = 0$.  It
is seen that the Bessel ansatz represents a rather good approximation
in the focusing case, but that the presence of the nonlinearity
changes the shape for small $r$, making it more peaked than the Bessel
profile.  In the defocusing case, the nonlinear solution instead has a
flatter form than the linear Bessel function,
Fig.~\ref{fig_var_num_cos_defocus}.  It is also seen that for
increasing $r_0$, or equivalently increasing $A_0$, the approximation
deteriorates.  This is due to the fact that in this case there exists
a threshold value for the amplitude in order to have well-behaved
solutions, cf.~(\ref{eq_limit}).  The critical value for the amplitude
is $A_0 = 1$.  The variational result also predicts this behaviour,
although, as is inferred from Eq.~(\ref{eq_var_result_amp}), the
critical value is found to be slightly different
\begin{equation}
A_0 > \sqrt{-\delta J_1^2(j_0)/(2 C_1 \kappa)} \approx 1.33.
\end{equation}
Clearly, as $A(0)$ approaches the threshold value, which is unity, we
expect the accuracy of the variational approximation to deteriorate.

\begin{figure}[htbp]
\begin{center}
\includegraphics[scale=0.6]{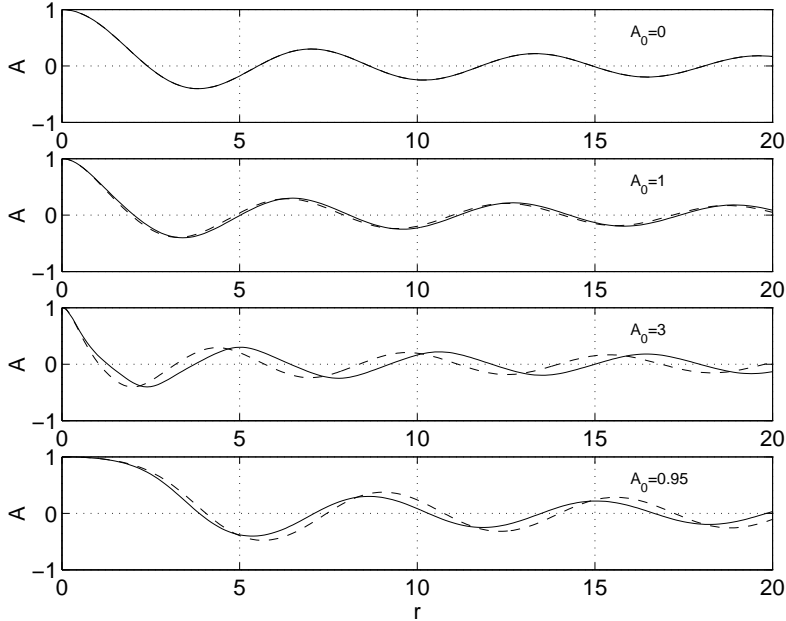}
\caption{The implicit analytical solution given by
Eq.~(\ref{integral}) (solid lines) together with the numerical result
(dashed lines). The different initial amplitudes are indicated in the
graph.  A defocusing nonlinearity is used in the fourth plot.}
\label{fig_global_approx}
\end{center}
\end{figure}

We now turn to an investigation of the overall behaviour of the
nonlinear Bessel beam profiles.  The fact that the main influence of
the nonlinear term is a rescaling of the radial coordinate makes it
reasonable to look for an approximate solution of the form
\begin{equation}
A(r) \approx A_0 J_0(f(r)),
\end{equation}
where $f(r)$ is a function of $r$, $\delta$, $\kappa$, and $A_0$.  It
is difficult to determine $f$ using analytical methods, but by
noticing that the linear solution can be written as
\begin{equation}
A = A_0 J_0 \left( \int_0^r \sqrt{\delta} \d r' \right),
\end{equation}
and by comparing with Eq.~(\ref{eq_nonlin_eq_sol_small_r}) it seems
reasonable that a good approximation should be obtained by the
implicit expression
\begin{equation}\label{integral}
A = A_0 J_0\left( \int_0^r \sqrt{\delta + \kappa A^2(r')} \d r'
\right).
\end{equation}
Although this is, in fact, an integral equation for $A(r)$, it
nevertheless provides a very simple formula for finding $A$
numerically. The corresponding approximate solution is compared to the
numerical solution of the full equation in
Fig.~\ref{fig_global_approx}. When the amplitude is low the two curves
are identical, since the ansatz then reduces to the Bessel function.
In the case of a focusing nonlinearity, there is good agreement
between the two approaches, but it is also seen that a phase shift
appears between the curves for increasing $A_0$. Quite good agreement
is seen also in the defocusing case.  In particular, the initial
flattening is well modelled.  The phase shift is now of the opposite
sign.

This approximate solution implies that the argument of the Bessel
function increases approximately as $\int_0^r\sqrt{\delta + \kappa
A^2} \d r'$, which is a nonlinear generalisation of the linear case.
Thus the main effect of a focusing nonlinearity is to increase the
curvature of the peaks by increasing the growth rate of the argument,
making the solution radially compressed.  In the defocusing case the
curvature is decreased, which is most clearly seen in the main lobe.

\begin{figure}[htbp]
\begin{center}
\includegraphics[scale=0.6]{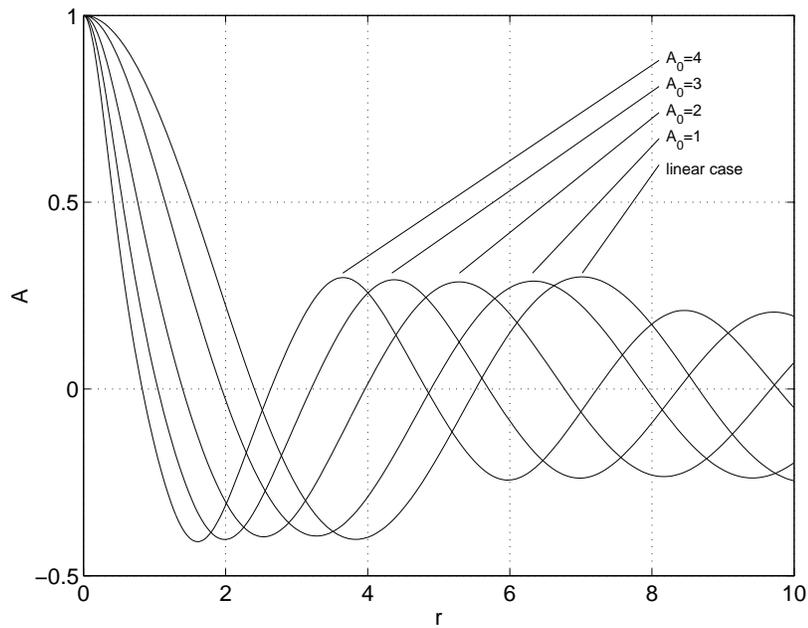}
\caption{A focusing nonlinear term gives rise to a radial compression,
which is illustrated using numerical simulations.}
\label{fig_num_focus}
\end{center}
\end{figure}

\begin{figure}[htbp]
\begin{center}
\includegraphics[scale=0.6]{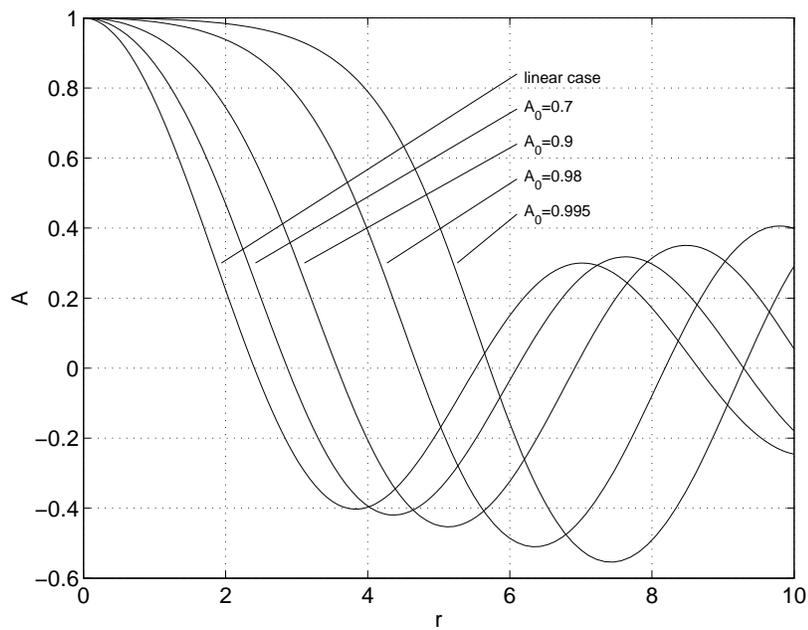}
\caption{Numerical solutions showing the influence from a defocusing
nonlinear term.} 
\label{fig_num_defocus}
\end{center}
\end{figure}

Finally, Figs.~\ref{fig_num_focus} and \ref{fig_num_defocus} further
illustrate the nonlinear deformations of the linear diffraction-less
Bessel solutions.  The numerically obtained curves clearly show the
features discussed above; the radial compression of the central lobe
in the focusing case and the radial expansion in the defocusing case.
The expansion effect in the latter case rapidly increases as the
amplitude approaches the critical value $A_0 = 1$, above which no
stationary solutions are possible.  The phase shifting effect of the
nonlinearity on the Bessel-like oscillations is also seen, the shift
changing sign with the sign of the nonlinearity.

\section{Analysis of nonlinear Bessel-Gauss beams}
The linear diffraction properties of Bessel-Gauss beams have been
analysed and solved analytically, see \cite{gori}.  In the present
section we will analyse the nonlinear dynamics of beams, which
initially have a profile in the form of a Bessel function truncated by
a Gaussian.  Since a general solution of this problem cannot be given,
we will use the virial theorem to obtain analytical information and
numerical simulations for determining the evolution of the beam
profile.

The virial theorem, see e.g.\ \cite{rasmussen,rypdal,fibich}, provides
exact and explicit information about the dynamic variation of the
width of the beam, $\sigma$, defined in the RMS sense as $\sigma^2
\equiv \langle r^2 \rangle$, where
\begin{equation}
\langle f(r) \rangle \equiv \frac{\int_0^\infty f(r) |\psi(z, r)|^2 r
\d r}{\int_0^\infty |\psi(z, r)|^2 r \d r}.
\end{equation}
The virial theorem asserts that
\begin{equation}
\frac{d^2 \sigma^2}{d z^2} = 8\frac{H}{I} = \mbox{constant},
\end{equation}
where $I$ and $H$ are invariants of the two-dimensional nonlinear
Schr\"odinger equation, Eq.~(\ref{eq_nlse}), and are defined as
follows
\begin{eqnarray}
I &=& \int_0^\infty |\psi(z, r)|^2 r \d r,\\
H &=& \int_0^\infty \left[ \left| \frac{\partial\psi(z, r)}{\partial
r} \right|^2 - \frac{\kappa}{2} |\psi(z, r)|^4 \right] r \d r.
\end{eqnarray}
This means the invariants correspond to the (integrated) beam
intensity and the Hamiltonian.  Thus, the virial theorem implies that
$\sigma^2$ must be a second order polynomial in $z$, with coefficients
determined by the initial beam profile, $\psi(0, r)$.  For initial
phase functions that do not depend on $r$, the linear term in $z$
vanishes, and the beam width varies as
\begin{equation}
\sigma^2(z) = \sigma^2(0) \left( 1 + \sign(H) \frac{z^2}{L_0^2}
\right)
\end{equation}
where $L_0$ is a characteristic length given by
\begin{equation}
L_0^{-2} = \left| \frac{4 H}{\sigma^2(0) I} \right| = \left| \frac{4
\int_0^\infty \left[ \left| \frac{\partial\psi(z, r)}{\partial r}
\right|^2 - \frac{\kappa}{2} |\psi(z, r)|^4 \right] r \d
r}{\int_0^\infty r^2 |\psi(0, r)|^2 r \d r} \right|.
\label{eq_L0}
\end{equation}
Clearly this approach cannot be used for analysing the linear, or the
nonlinearly modified, stationary Bessel beam solutions of the
nonlinear Schr\"odinger equation since all integrals involved in the
virial theorem are infinite. However, for a physical beam, with finite
integral content, the virial theorem is useful.  In general it is seen
that with weak nonlinear focusing effects, the Hamiltonian is positive
and the RMS width will increase quadratically with a characteristic
diffraction length given by $L_0$. When the amplitude of the beam
increases, the Hamiltonian decreases and eventually changes sign.
This implies that the RMS width goes to zero after a finite length
equal to $L_0$---the well known nonlinear collapse phenomenon, where
$L_0$ now plays the role of collapse length.

For Bessel-Gauss beams, \cite{gori}, the initial profile is of the
form
\begin{equation}
\psi(0, r) = A_0 J_0 \left( \frac{r}{r_0} \right) \exp
\left(-\frac{r^2}{2 \rho_0^2} \right).
\end{equation}
Inserting this into Eq.~(\ref{eq_L0}) we obtain the following
expression for the characteristic length $L_0$
\begin{equation}
\left( \frac{r_0}{2 L_0} \right)^2 = \frac{S_1(\mu) - \Lambda
S_2(\mu)}{S_3(\mu)}
\end{equation}
where $\mu = r_0^2/\rho_0^2$, $\Lambda = \kappa A_0^2 r_0^2$, and the
integrals $S_n$, $n = 1, 2, 3$, are given by
\begin{eqnarray}
S_1(\mu) &=& \int_0^\infty [\mu x J_0(x) + J_1(x)]^2 \exp(-\mu x^2) x
\d x, \\
S_2(\mu) &=& \int_0^\infty J_0^4(x) \exp(-2 \mu x^2) x \d x, \\
S_3(\mu) &=& \int_0^\infty x^2 J_0^2(x) \exp(-\mu x^2) x \d x.
\end{eqnarray}
Since we are primarily interested in the case when the Gauss function
truncates the outer parts of the Bessel function, we have $\mu =
r_0^2/\rho_0^2 \ll 1$. In this limit, the asymptotic values of the
integrals, $S_n$, $n = 1, 2, 3$, are obtained analytically as
\begin{eqnarray}
S_1(\mu) &\approx& \frac{1}{2\sqrt{\pi}} \frac{1}{\sqrt{\mu}}, \\
S_2(\mu) &\approx& D_1- \frac{3}{4 \pi^2} \ln \mu, \\
S_3(\mu) &\approx& \frac{1}{4 \sqrt{\pi}} \frac{1}{\mu^{3/2}},
\end{eqnarray}
Since $S_2(\mu)$ goes rather slowly towards infinity as $\mu $ becomes
small, it is necessary to determine the constant $D_1$ in order to
have good accuracy for finite $\mu$. Using numerical evaluation of the
integral we find $D_1 \approx 0.202$. This implies that the
characteristic length can be approximated as
\begin{equation}
\left( \frac{r_0}{2 L_0} \right)^2 \approx 2 \mu \left[ 1 - \Lambda
  \sqrt{\mu} \left( D_2 - \frac{3 \ln \mu}{2 \pi^{3/2}} \right)
  \right],
\label{eq_virial_diffraction_length}
\end{equation}
with $D_2 \approx 0.715$.  In the linear case, the characteristic
length is seen to scale simply as $L_0 \propto \rho_0$, i.e., the
diffraction is determined solely by the truncation radius and as
$\rho_0 \to \infty$, the non-diffracting Bessel beam is recovered. For
increasing values of the nonlinearity parameter, $\Lambda$, but for a
fixed truncation radius, the value of $L_0$ increases and for a
certain critical value of $\Lambda$, the nonlinearity balances the
diffraction to give a Bessel-Gauss beam, which is diffraction-less in
the RMS sense.

\section{Dynamics of Bessel-Gauss beams}
\begin{figure}[htbp]
\begin{center}
\includegraphics[scale=0.6]{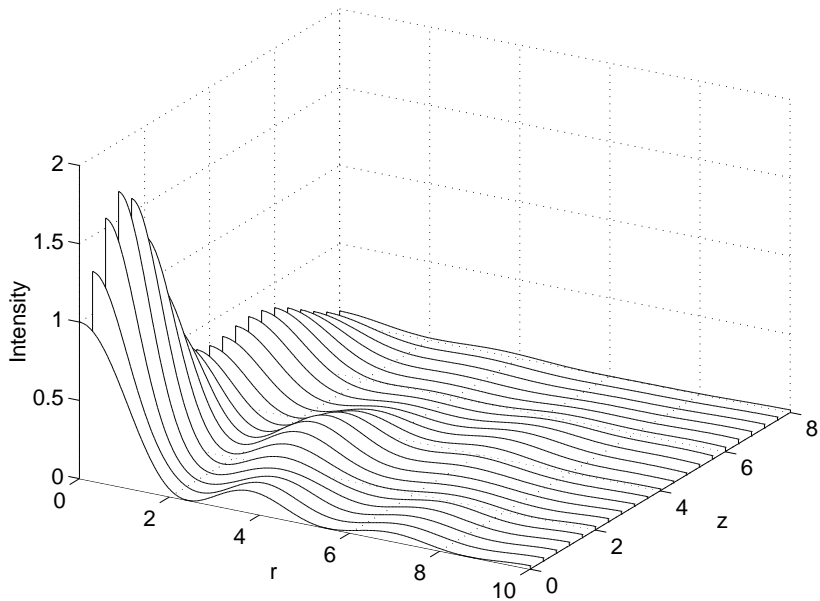}
\caption{An example of the evolution of the radial intensity profile
with distance of propagation.} 
\label{fig_propagation_3D}
\end{center}
\end{figure}

\begin{figure}[htbp]
\begin{center}
\includegraphics[scale=0.6]{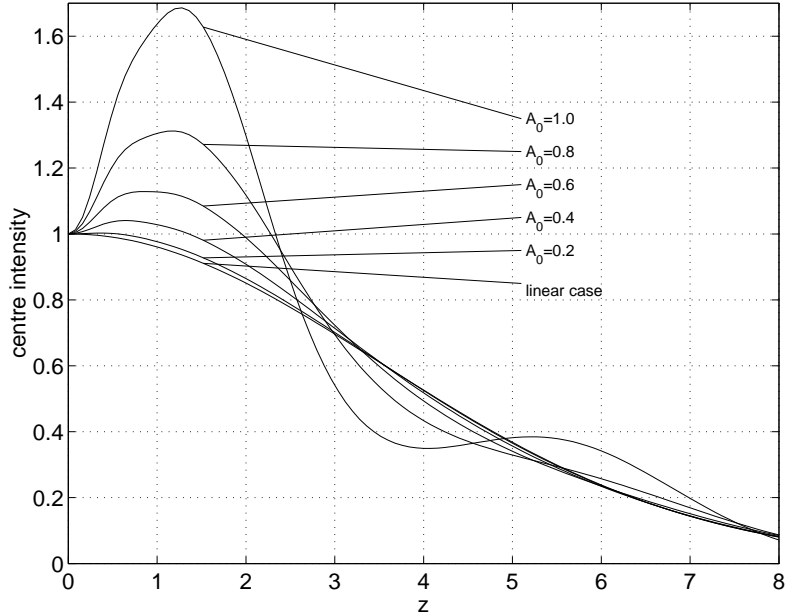}
\caption{The intensity at $r = 0$ as a function of the propagation
distance using different initial amplitudes.}
\label{fig_centre_intensity}
\end{center}
\end{figure}

When a truncated linear Bessel-Gauss beam propagates, the diffraction
initially affects only the outermost part of the pulse, where the
truncation has changed it from the Bessel shape.  The central parts
are initially diffraction balanced and remains so until the
``diffraction front'' propagating inwards from the outer parts
eventually reach the inner lobes and also these parts start to
diffract outwards.  On the other hand, the nonlinear effect is
strongest at the centre of the beam, where the intensity is highest,
and with a focusing nonlinear term, the main lobe will start to
compress.  In fact, it will start to compress irrespective of the
degree of nonlinearity since the linear diffraction is already
balanced.  If the nonlinear effect is weak the compression will
eventually stop, and diffraction will become the dominating effect.
This evolution is illustrated in Fig.~\ref{fig_propagation_3D}, where
an FDTD simulation using $A_0 = 1$ and $\mu = 0.01$ is shown.
Although the central parts initially compress, the virial theorem
predicts beam broadening in the RMS sense.  Clearly this is no
contradiction since the broadening of the outer parts more than
compensate the compression of the centre.  In order to further
illustrate how the main lobe is compressed, the intensity at $r = 0$
has been plotted for different initial amplitudes as a function of
propagation distance in Fig.~\ref{fig_centre_intensity}.  The curves
have been normalised with respect to their amplitudes at $r = 0$ and
$\mu = 0.01$.  We emphasise the oscillating behaviour for the highest
amplitudes.  If the nonlinear effect is sufficiently strong, the
virial theorem predicts a collapse of the beam to zero RMS width.
According to Eq.~(\ref{eq_virial_diffraction_length}), the amplitude
threshold for this type of behaviour is $A_0 >A_c \approx 2.26$ for
$\mu=0.01$.  It is well known that the nonlinear evolution of
two-dimensional beams may lead to a break up of the beam into a
diffracting background profile with a monotonously compressing
filament, that collapses in a finite distance of propagation.  Thus,
the width of the filament goes to zero and the intensity becomes
infinite whereas the beam width in the RMS sense still increases.  The
simulations for the present case of Bessel-Gauss beams show that when
the amplitude is increased further above $A_0 = 1$, the small second
peak of Fig.~\ref{fig_centre_intensity} will start to dominate and
eventually the simulations indicate a collapse of the second peak,
although the RMS width still tends to infinity.  In fact, even if the
RMS width remains constant, the beam should still be able to undergo
partial collapse.

Much effort has been devoted to the study of two-dimensional collapse
phenomena induced by the Kerr nonlinearity, see
e.g.~\cite{rasmussen,rypdal,fibich} and references therein. In
particular, it has been found that the virial theorem poses a
sufficient but not necessary condition for the occurrence of a
singularity where the amplitude becomes infinite.  Thus the appearance
of a partial collapse singularity below the threshold for a global
collapse, as predicted by the virial theorem, is in accordance with
earlier results.

\section{Conclusions}
Based on the nonlinear Schr\"odinger equation in cylindrical geometry
we have studied the modification of the diffraction-less linear Bessel
beams caused by the nonlinear Kerr effect.  The stationary as well as
the dynamic properties of the solutions to this equation have been
analysed and both analytical and numerical techniques have been used.
The investigation shows that the nonlinearity primarily affects the
main and inner lobes of the Bessel beams.  In the case of the
stationary solutions, the central region of the nonlinear Bessel beam
tends to become radially more narrow or more extended depending on
whether the nonlinearity is focusing or defocusing, respectively.
Asymptotically the solutions are still of the same oscillating form as
the diffraction-less Bessel beams, the only remaining feature of the
nonlinearity being a phase shift as compared to the linear case.
However, in the case of the defocusing nonlinearity, there is a finite
amplitude threshold for well-behaved solutions to exist.  Above this
limit the nonlinear diffraction effect becomes larger than the linear
effect, which counteracts the diffraction, and no solutions are
possible which vanish at infinity.

The properties of Gaussian-truncated Bessel beams have also been
studied in the presence of the Kerr nonlinearity.  It has been shown,
using the virial theorem, that a non-diffracting situation in the RMS
sense is possible to obtain by balancing nonlinear focusing and linear
diffraction.  However, this situation does not correspond to a
stationary case of the beam profile.  Significant redistribution of
the beam occurs and using numerical simulations, the dynamic
interaction between linear diffraction and nonlinear focusing has been
analysed for varying degrees of nonlinearity.  It has been found that,
in particular the central parts of the beam may become significantly
distorted and may even partially collapse even though the beam width,
defined in the RMS sense, remains constant or even increases.  This
result is in agreement with the classical picture of dynamic self
focusing of two-dimensional beams in nonlinear Kerr media.

\end{document}